\documentclass[aps,pra,showpacs,notitlepage]{revtex4-1}

\usepackage{graphicx}

\begin{document}

\title{Quantum interference of position and momentum:\\
a particle propagation paradox}

\author{Holger F. Hofmann}
\email{hofmann@hiroshima-u.ac.jp}
\affiliation{
Graduate School of Advanced Sciences of Matter, Hiroshima University,
Kagamiyama 1-3-1, Higashi Hiroshima 739-8530, Japan
}

\begin{abstract}
Optimal simultaneous control of position and momentum can be achieved by maximizing the probabilities of finding their experimentally observed values within two well-defined intervals. The assumption that particles move along straight lines in free space can then be tested by deriving a lower limit for the probability of finding the particle in a corresponding spatial interval at any intermediate time $t$. Here, it is shown that this lower limit can be violated by quantum superpositions of states confined within the respective position and momentum intervals. These violations of the particle propagation inequality show that quantum mechanics changes the laws of motion at a fundamental level, providing a new perspective on causality relations and time evolution in quantum mechanics.
\end{abstract}

\maketitle

Despite its great success in explaining a wide range of phenomena, quantum mechanics fails to provide a convincing explanation of how particles move in free space. In the standard formalism, the quantum state of a particle is represented by a wavefunction, and this wavefunction spreads out in space as a particle propagates. The spatial distribution of the wavefunction represents the probability of finding the particle at a given position, assuming that it is detected at the appropriate time $t$. However, the wavefunction at time $t$ does not convey any information about the position of the particle at any other time. There have been a number of approaches that attempt to identify elementary trajectories which might explain the statistical patterns of quantum dynamics \cite{Koc11,Hal14,Mag16,Zho17}, but all of them suffer from untraceable additional assumptions, and there remains a great deal of ambiguity in the interpretation of quantum statistics in space and time \cite{Lun11,Sch13,Bli13,Mor16}. Nevertheless, there is some evidence that the evolution of wavefunctions in time and space is not consistent with the assumption of a constant velocity determined by the momentum. In particular, it has been shown that the current of probability described by the Schr\"odinger equation can point in the opposite direction of all available momentum components \cite{Bra94,Hal13a}. It is therefore reasonable to ask whether quantum theory might require modifications to Newton's first law \cite{War02,Pfi04,Hal13b}. Unfortunately, it is difficult to identify such modifications using only a statistical analysis of the available data. As pointed out by Schleich et al. \cite{Sch13}, the Wigner function can actually be interpreted as a reconstructed probability distribution over straight line trajectories defined by initial position and momentum. The deviations from classical physics caused by quantum interference then result in negative values of the quasi-probability expressed by the Wigner function. Similarly, the measurement of the wavefunction presented by Lundeen et al. \cite{Lun11} characterize quantum interferences in terms of the complex-valued quasi-probability of position and momentum expressed by the Dirac distribution \cite{Hof12}. In the light of these unresolved ambiguities, it is unclear whether the attribution of trajectories has any physical meaning at all. In fact, a number of quantum paradoxes suggest that the assignment of intermediate positions to moving particles is highly problematic \cite{box,Har92,Res04,Yok09,Lun09,Aha13,Den14}. However, none of these paradoxes have directly addressed the question of particle motion itself. 

In the present paper, I investigate the paradoxical aspects of particle motion in free space by formulating the statistical limit that Newton's first law imposes on the separate statistics of initial position, momentum, and an intermediate position at a specific time $t$. The argument relies on the observation that quantum interference between a quantum state confined in a position interval and a quantum state confined in a momentum interval can result in a probability of more than 50 \% of finding the particle in the two intervals, which seems to require a minimal probability of finding the particle in both intervals at the same time and therefore puts a limit on the possible straight line trajectories within both intervals. However, the quantum interference effect at intermediate times results in much lower probabilities for the positions of these hypothetical straight line trajectories, resulting in a violation of the statistical limit for Newton's first law. In the following, this violation of Newton's first law is demonstrated by choosing a specific superpostion of a state localized in a position interval $L$ and a state localized in a momentum interval $B$. It is then shown that this superposition can be optimized by varying the width of the position and momentum interval, resulting in a violation of the statistical limit for Newton's first law by about 8 \%. This specific example demonstrates the possibility of obtaining clear and unambiguous experimental evidence that quantum particles do not move along straight lines in free space, indicating that quantum mechanics requires a fundamental revision of Newton's first law in order to explain the effects of quantum interferences between position and momentum. 

To put the problem into perspective, it may be good to remember that Heisenberg's formulation of dynamics actually suggests the continuing validity of Newton's first law in the form of the operator equation
\begin{equation}
\label{eq:Heis}
\hat{x}(t) = \hat{x}(0) + \frac{1}{m} \hat{p} \; t.
\end{equation}
In quantum mechanics, this relation fully defines the time evolution for particles in free space, where no forces act on the moving body and the momentum $\hat{p}$ is conserved at all times. In particular, it is possible to identify the momentum $\hat{p}$ with positions at $t \to \infty$ if a reasonable limit can be given for the possible values of $\hat{x}(0)$. However, it remains unclear how the operators relate to specific values of position and momentum since they do not commute with each other and have no joint eigenstates. To properly address the problem of motion in quantum mechanics, we must therefore identify experimentally observable evidence for the relation between positions at different times, even though the uncertainty relations between position and momentum means that this evidence can only come in the form of statistical estimates based on incomplete information about the initial position and momentum at $t=0$. A good way to do this is to use the probability $P(L)$ of finding the initial position in an interval of width $L$ around a central value of $x(0)=0$. Likewise, the momentum distribution can be characterized by the probability $P(B)$ of finding the momentum in an interval of width $B$ around a central value of $p=0$. For particles with positions in the interval $L$ and momenta in the interval $B$, motion in a straight line requires that the intermediate positions are found in an interval given by
\begin{equation}
\label{eq:limit}
|x(t_M)| \leq \frac{1}{2} L + \frac{1}{2m} B \; t_M.
\end{equation}
This interval has a width of $M=L+B\,t_M/m$. Newton's first law requires that particles in the interval $L$ at $t=0$ and {\it also} in the interval $B t/m$ at $t \to \infty$ must move through the interval $M$ at time $t=t_M$. Statistically, these conditions are satisfied with a probability of $P(L \;\mbox{AND}\; B)$ that can be related to the experimentally observable probabilities of finding the particle in $L$ or in $B$ by
\begin{equation}
\label{eq:joint}
P(L\; \mbox{AND}\; B) = P(L) + P(B) - P(L\; \mbox{OR}\; B).
\end{equation}
The lowest possible value of $P(L\; \mbox{AND}\; B)$ is obtained when $P(L\; \mbox{OR}\; B)=1$, which means that all particles not found in $L$ are found in $B$ and vice versa. Since Eq.(\ref{eq:limit}) requires that $P(M) \geq P(L\; \mbox{AND}\; B)$, the probability $P(M)$ must satisfy the inequality
\begin{equation}
\label{eq:propcond}
P(M) \geq P(L) + P(B) - 1.
\end{equation}
This propagation inequality defines a fundamental statistical limit for the experimentally observable probability densities of particles moving along a straight line. Specifically, the probabilities $P(L)$, $P(B)$ and $P(M)$ are given by integrals of the spatial probability densities for $t=0$, for $t \to \infty$, and for an intermediate time $t=t_M$, all of which can be measured in separate quantum measurements. It is therefore possible to test the validity or the violation of the inequality for any quantum state of a particle propagating in free space by performing three separate position measurements. Eq.(\ref{eq:propcond}) thus provides us with an unambiguous experimental criterion for the failure of Newton's first law in quantum mechanics.

To achieve a sufficiently strong violation of Eq. (\ref{eq:propcond}), the initial state $\mid \psi \rangle$ needs to be confined simultaneously in position and in momentum to achieve a probability sum $P(L)+P(B)$ considerably larger than one. Particularly high values for this probability sum can be obtained by equal superpositions of a state $\mid L \rangle$ localized in the spatial interval $L$ around $x(0)=0$ and a state $\mid B \rangle$ localized in the momentum interval $B$ around $p=0$, with constructive interferences enhancing the probability sum $P(L)+P(B)$ well beyond the values achieved by a mixture of the states. In general, it is possible to consider states with a non-vanishing probability of finding the particle outside of the respective intervals, as long as these probabilities are negligibly small when compared to the probabilities added by quantum interference. However, it seems that the most natural experimental implementation of a state $\mid L \rangle$ confined to a position interval $L$ is the passage of a particle through a slit of that width at time $t=0$. Likewise, $\mid B \rangle$ can be implemented by passage through a ``momentum slit'' filter of width $B$. In the position basis at $t=0$, the wavefunction of this momentum state $\mid B \rangle$ reads
\begin{equation}
\label{eq:Bshape}
\langle x \mid B \rangle = \sqrt{\frac{B}{2 \pi \hbar}} \left(\frac{2 \hbar}{B x} \sin\left(\frac{B}{2 \hbar} x \right)\right). 
\end{equation}
The inner product of this momentum state with the position state $\mid L \rangle$ is given by the sine integral
\begin{equation}
\label{eq:overlap}
\langle L \mid B \rangle = \sqrt{\frac{B L}{2 \pi \hbar}} \left(\frac{4 \hbar}{BL} \mbox{Si}\left(\frac{B L}{4 \hbar}\right) \right).
\end{equation}
Thus the overlap only depends on the ratio of $B L/\hbar$, which evaluates the suppression of uncertainty by simultaneous confinement of a particle in the intervals $L$ and $B$. For $B L < \hbar$, the sine integral can be approximated by $B L/ (4 \hbar)$ and the expression in parenthesis is approximately equal to one. The inner product $\langle L \mid B \rangle$ is then given by the square root of the ratio $BL/(2 \pi \hbar)$.

As mentioned above, the probability $P(L)+P(B)$ is enhanced by the superposition of $\mid L \rangle$ and $\mid B \rangle$, with a maximal probability obtained from an equal superposition of
\begin{equation}
\mid \psi \rangle = \frac{1}{\sqrt{2(1+\langle L \mid B \rangle)}}
\left(\mid L \rangle + \mid B \rangle \right).
\end{equation}
For this state, the probabilities $P(L)$ and $P(B)$ are
\begin{equation}
P(L)=P(B)=\frac{1}{2}\left(1 + \langle L \mid B \rangle \right).
\end{equation}
The particle propagation inequality thus requires a minimal probability $P(M)$ of finding the particle in an intermediate interval $M$ around $x(t_M)=0$ given by the overlap of the quantum state components $\mid L \rangle$ and $\mid B \rangle$,
\begin{equation}
P(M) \geq \langle L \mid B \rangle.
\end{equation}
This limit can now be compared to the actual probability densities observed at various times $t$. Here, we can identify the essential difference between a joint definition of position and momentum and a superposition of the position state $\mid L \rangle$ and the momentum state $\mid B \rangle$. Instead of enhancing only the probabilities of results within the interval $M$, interference is equally distributed over the complete region of spatial overlap between $\langle x \mid \hat{U}(t) \mid L \rangle$ and $\langle x \mid \hat{U}(t) \mid B \rangle$. Specifically, the slit function $\langle x \mid L \rangle$ changes to a sinc function for $t > m L^2/(2 \pi \hbar)$, with
\begin{equation}
\label{eq:Lshape}
\langle x \mid \hat{U}(t) \mid L \rangle = \sqrt{\frac{m L}{2 \pi \hbar t}} \left( \frac{2 \hbar t}{m L x} \sin\left(\frac{m L}{2 \hbar t} x \right)\right) \exp\left( i \frac{m}{2 \hbar t} x^2 - i\frac{\pi}{4} \right). 
\end{equation}
The shape of this wavefunction corresponds to that of the momentum state given in Eq.(\ref{eq:Bshape}), except for the complex phase factor and the time dependent width. The distribution shown in Eq.(\ref{eq:Lshape}) describes the effects of the momentum distribution of $\mid L \rangle$, which dominates the dynamics for times greater than $m L^2/(2 \pi \hbar)$. Likewise, Eq.(\ref{eq:Bshape}) describes the position distribution of $\mid B \rangle$, which dominates the dynamics for times smaller than $2 \pi \hbar m/B^2$. At $t_M=m L/B$, the wavefunctions are perfectly matched, resulting in the minimal value of $P(M)$ as both quantum state components contribute equally to the probability density near $x(t_M)=0$. The complete probability distribution at $t_M=m L/B$ is given by
\begin{equation}
\label{eq:pattern}
|\langle x \mid \hat{U}\left(\frac{mL}{B}\right) \mid \psi \rangle|^2
= \frac{2 (\langle L \mid B \rangle)^2}{(1+\langle L \mid B \rangle) L} \left(\frac{2 \hbar}{B x} \sin\left(\frac{B x}{2 \hbar}\right) \right)^2 \left(\cos\left(\frac{\pi}{2 (\langle L \mid B \rangle)^2} \left(\frac{B x}{2 \hbar}\right)^2 - \frac{\pi}{8} \right)\right)^2.
\end{equation}
Note that the inner product $\langle L \mid B \rangle$ given in Eq.(\ref{eq:overlap}) serves here as a convenient expression for the uncertainty product $B L$, which is the only free parameter that still needs to be optimized. We can perform the optimization by maximizing the defect probability that quantifies the violation of the propagation inequality Eq.(\ref{eq:propcond}),
\begin{equation}
P_{\mathrm{defect}} = P(L)+P(B)-1-P(M).
\end{equation}
A reliable estimate of the defect probability can be obtained by using the envelope function of the interference pattern in Eq.(\ref{eq:pattern}), multiplying the maximal possible density with the width $M=2 L$ of the interval in which particles with $|x(0)|<L/2$ and $|p|<B/2$ would have to be found. The result is a lower limit for $P_{\mathrm{defect}}$ of
\begin{equation}
P_{\mathrm{defect}} \geq \langle L \mid B \rangle - \frac{4 (\langle L \mid B \rangle)^2}{1+\langle L \mid B \rangle}.
\end{equation} 
\begin{figure}[th]
\vspace{-6cm}
\begin{picture}(500,500)
\put(0,0){\makebox(480,450){
\scalebox{0.85}[0.85]{
\includegraphics{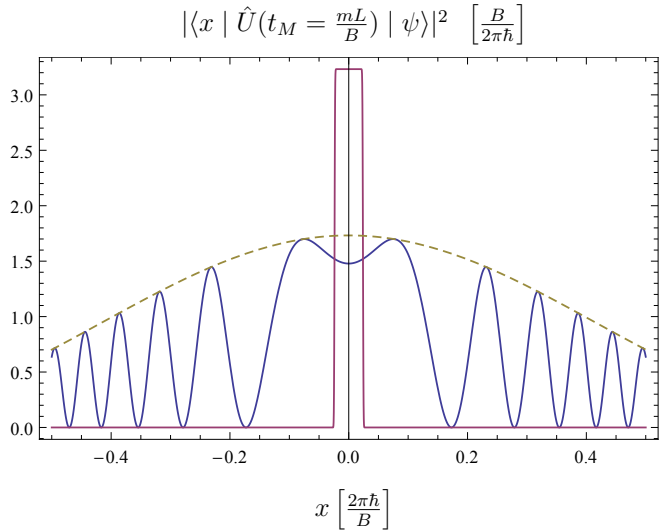}}}}
\end{picture}
\vspace{-4cm}
\caption{\label{result}
Probability density of position observed at $t_M=mL/B$ for an optimized overlap of $\langle L \mid B \rangle = 2/\sqrt{3}-1$ ($B L \approx 0.024 (2 \pi \hbar)$). The dotted line marks the envelope function of the interference pattern used in the optimization of the inequality violation. The rectangular line marks the interval $M=2 L$ ($|x|\leq L$) and the probability density corresponding to the minimal value of $P(M)$ needed to satisfy the propagation inequality Eq. (\ref{eq:propcond}). The observation of probability densities at less than half the value needed to satisfy the propagation inequality demonstrates its violation by the quantum state $\mid \psi \rangle$, with the defect probability $P_{\mathrm{defect}}$ given by the area between the rectangular line and the experimentally observed probability density. 
}
\end{figure}
Note that this equation describes a trade-off between the probability sum $P(L)+P(B)$ and the probability density in the intermediate interval $M$, where the positive effects of increasing overlap is graduately offset by the associated increase in $P(M)$. The maximal value of $P_{\mathrm{defect}}$ is obtained for $\langle L \mid B \rangle =2/\sqrt{3}-1$, which corresponds to an uncertainty product of $B L \approx 0.024 (2 \pi \hbar)$. Fig. \ref{result} shows a comparison of the experimentally observable probability distribution at $t=m L/B$ for the optimized product of $B$ and $L$. The minimal probability that needs to be localized within the interval of $M=2 L$ for propagation in a straight line between $x(0)$ and $x(\to \infty)$ is shown by a rectangular distribution. As can be seen in the figure, the actual probability density is less than half the minimal probability density that would be required to satisfy the propagation inequality Eq.(\ref{eq:propcond}). The specific value obtained from the probabilities $P(L)$ and $P(B)$ for the statistical limit is
\begin{equation}
P(L)+P(B)-1 = 0.155,
\end{equation}
which means that the probability under the rectangular distribution in Fig. \ref{result} is about 15.5\%. The estimate for $P(M)$ obtained from the envelop function shown as a dotted line in Fig. \ref{result} is
\begin{equation}
P(M) \leq 0.083.
\end{equation}
Clearly, the percentage of particles found in the interval $M$ is much lower than the minimal percentage required for particles propagating in accordance with Newton's first law. The result can also be given in terms of the defect probability, which quantifies the violation of the propagation inequality Eq.(\ref{eq:propcond}) in terms of the minimal percentage of particles found in interval $B$ and $L$, but not in interval $M$. Based in the envelope function of the interference pattern, its minimal value is
\begin{equation}
P_{\mathrm{defect}} \geq 0.072.
\end{equation}
The actual defect probability is even higher, due to the phase shift of $\pi/8$ in the interference pattern at $x(t)=0$. A rough estimate indicates a value near 8\%, more than half of the limit given by the propagation inequality Eq.(\ref{eq:propcond}). 

The particle propagation paradox outlined above applies equally to all particles moving in free space. Experimentally, it requires (a) the ability to prepare a superposition of states $\mid L \rangle$ and $\mid B \rangle$, (b) a measurement of $\hat{x}$ at $t=0$ and at $t=t_M$, and (c) a measurement of $\hat{p}$, possibly realized by a measurement of $\hat{x}$ at sufficiently large values of $t$. Note that the control of propagation time is essential in all three experimental procedures. Since we are mostly interested in one-dimensional motion, the easiest solution is to use particles propagating at a well-defined velocity of $c$ along the $z$-axis, so that the distance along $z=c t$ can be used as a measure of propagation time. Specifically, this method is commonly applied to photons, where the paraxial approximation results in an effective mass of $m=2\pi \hbar/(c \lambda)$ for the transverse dynamics of photons of wavelength $\lambda$. The states $\mid L \rangle$ and $\mid B \rangle$ can then be prepared by passing the photons through slits at $z=0$ or at $z \to \infty$, where lenses can be used to adjust the effective distance. It should be noted that this procedure has already been realized as a method of preparing spatial qubits, and the results indicate that rectangular wavefunctions can indeed be produced experimentally with very high fidelity \cite{Nev07,Tag08,Pee09,Sol11}. Superpositions of different state preparations can be realized in this system by performing the appropriate operations in different arms of a two path interferometer. Thus photons seem to be the ideal system for an experimental realization, especially for the quantum states discussed above. 

It should also be noted that the states used to demonstrate the violation of the particle propagation inequality are by no means unique. While the present selection seems to be most promising because of the comparatively high values of $P(L)+P(B)$, the result of $1.155$ for the optimized product of $L$ and $B$ suggests that a wide range of other states with similar distributions of position and momentum can also achieve 
$P(L)+P(B)>1$. In particular, the probabilities of finding the particle in the intervals $L$ and $B$ do not need to be one for $\mid L \rangle$ and $\mid B \rangle$, as long as constructive interference between the states makes up for the difference. The analysis above suggests that it is the quantum interference between a state confined in the spatial interval $L$ with a state confined in the momentum interval $B$ that causes the violation of the particle propagation inequality, where the product $L B$ of the confinement must be below the uncertainty limit of $2 \pi \hbar$, but sufficiently high to obtain a significant interference effect. Although a precise analysis is beyond the scope of the present paper, it seems reasonable to expect that a wide variety of states can satisfy these criteria.

In conclusion, superpositions of rectangular position states $\mid L \rangle$ and momentum states $\mid B \rangle$ result in violations of the statistical limits imposed by the assumption that particles move along straight lines according to Newton's first law. By optimizing the uncertainty product $B L$ of the widths in position and momentum, a maximal probability defect of more than seven percent is obtained, which is a factor of two below the limit given by the inequality. The propagation inequality Eq.(\ref{eq:propcond}) thus provides a quantitative and experimentally testable criterion for the deviation of quantum dynamics from the classical laws of motion. Similar to other inequality violations, the violation of the propagation inequality does not provide any evidence in favor or against any of the possible interpretations of dynamics in quantum mechanics. However, it should be recognized that the standard formalism of quantum mechanics describes particle propagation as a quantum coherent process, and the propagation inequality shows just how much this quantum coherent process can deviate from Newton's first law. As shown by Eq.(\ref{eq:Heis}), the key to this inequality violation lies in the different roles of operators and eigenvalues \cite{Nii17}. By analyzing the actual measurement statistics of the three non-commuting observables $x(t)$, $x(0)$ and $p$, we can show that no simultaneous assignment of eigenvalues to the three operators can satisfy Eq.(\ref{eq:Heis}). Instead, the relations between non-commuting properties are defined by quantum interference effects, as shown by the quantum statistics of $x(t)$ given by Eq.(\ref{eq:pattern}) and shown in Fig. \ref{result}. Incidentally, this interference pattern is quite similar to the relation between $x(0)$ and $p$ observed in weak measurements and described by the Dirac distribution \cite{Lun11,Hof12}. It may therefore be possible to explain both the non-classical features of quantum statistics and the quantum modifications to the classical laws of motion by taking a closer look at the fundamental role of quantum coherent phases in the dynamics of quantum systems \cite{Hof16,Hib17}.

The propagation paradox presented above provides an unambiguous demonstration that classical notions of causality break down in the limit of precisely defined quantum statistics. By using statistical arguments, it can be shown that the uncertainty principle is not sufficient to explain the observation of quantum interference effects in free space. Clearly, quantum mechanics replaces Newton's first law with a different form of causality, and the maximal violation of the propagation inequality by a superposition of position and momentum could represent an important first step towards a more thorough investigation of this relation between quantum interference effects and the laws of motion.

\vspace{0.2cm}

This work has been supported by CREST, Japan Science and Technology Agency.

\end{document}